\documentclass[journal]{IEEEtran}

\title{Wireless Information Surveillance via STAR-RIS}

\author{\IEEEauthorblockN{Fatemeh Jafarian$^\dag$, Mehrdad Ardebilipour$^\dag$, Mohammadali Mohammadi$^\ddag$, and Michail Matthaiou$^\ddag$}\\
\small{
 $^\dag$ Department of Electrical Engineering, K. N. Toosi University
of Technology, Tehran, Iran\\
$^\ddag$Centre for Wireless Innovation (CWI), Queen's University Belfast, U.K.\\
Email:\{fatemeh.jafarian@email.kntu.ac.ir, mehrdad@eetd.kntu.ac.ir\}, \{m.mohammadi, m.matthaiou}@qub.ac.uk\}
}

\usepackage{xcolor,soul,framed} 
\colorlet{shadecolor}{yellow}
\usepackage[pdftex]{graphicx}
\graphicspath{{../pdf/}{../jpeg/}}
\DeclareGraphicsExtensions{.pdf,.jpeg,.png}

\usepackage{amsmath,epsfig,amssymb,verbatim,amsopn,cite,multirow}
\usepackage{amsthm}
\usepackage{balance}
\usepackage[short]{optidef}
\usepackage{algpseudocode}
\usepackage[all]{xy}  
\usepackage{url}
\usepackage{amsfonts}
\usepackage{amssymb}
\usepackage{epsfig}
\usepackage{epstopdf}
\usepackage{bm}
\usepackage{balance}
\usepackage{graphicx}
\usepackage{subcaption}
\usepackage{footnote}
\usepackage[norelsize, linesnumbered, ruled, lined, boxed, commentsnumbered]{algorithm2e}

\usepackage[nodisplayskipstretch]{setspace}
\newcommand{\qa}{{\bf a}}

\newcommand{\qf}{{\bf f}}
\newcommand{\qg}{{\bf g}}
\newcommand{\qh}{{\bf h}}

\newcommand{\qn}{{\bf n}}

\newcommand{\qw}{{\bf w}}

\newcommand{\qA}{{\bf A}}
\newcommand{\qB}{{\bf B}}

\newcommand{\qH}{{\bf H}}
\newcommand{\qI}{{\bf I}}

\newcommand{\qQ}{{\bf Q}}

\newcommand{\qW}{{\bf W}}

\newcommand{\diag}{\mathrm{diag}}
\newcommand{\RD}{\mathtt{RD}}
\newcommand{\EE}{\mathtt{EE}}
\newcommand{\RE}{\mathtt{RE}}
\newcommand{\SR}{\mathtt{SR}}
\newcommand{\ER}{\mathtt{ER}}
\newcommand{\MRT}{\mathtt{MRT}}
\newcommand{\MRC}{\mathtt{MRC}}
\newcommand{\ZF}{\mathtt{ZF}}
\newcommand{\NOP}{P_{\mathrm{NOP}}}
\newcommand{\Ps}{P_\mathtt{s}}

\newcommand{\roE}{P_{\mathtt{E}}}
\newcommand{\SnD}{\sigma_{\mathtt{D}}^{2}}
\newcommand{\SnE}{\sigma_{\mathtt{E}}^{2}}

\newcommand{\SINRE}{\mathrm{SINR}_{\mathtt{E}}}
\newcommand{\SINRD}{\mathrm{SINR}_{\mathtt{D}}}

\newcommand{\Ex}{\mathbb{E}}
\newcommand{\trace}{\mathrm{tr}}
\newcommand{\Diag}{\mathrm{diag}}
\newcommand{\Rank}{\mathrm{Rank}}

\newcommand{\Ev}{\text{E}}
\newcommand{\ST}{\text{ST}}
\newcommand{\SD}{\text{SR}}
\newcommand{\cs}{\mathtt{bs}}
\newcommand{\RZF}{\mathtt{RZF}}
\newcommand{\TZF}{\mathtt{TZF}}

\newcommand{\hRD}{\qh_{\RD}}
\newcommand{\hSR}{\qh_{\SR}}
\newcommand{\HHRD}{\qH_{\RD}}
\newcommand{\HHEE}{\qH_{\EE}}
\newcommand{\tilHHEE}{\tilde{\qH}_{\EE}}
\newcommand{\HHRE}{\qH_{\RE}}

\newcommand{\HHER}{\qH_{\ER}}
\newcommand{\wt}{\qw_{t}}
\newcommand{\wrr}{\qw_{r}}

\newcommand{\Otetr}{\overline{\boldsymbol{\theta}}_r}
\newcommand{\Otett}{\overline{\boldsymbol{\theta}}_t}
\newcommand{\Wtetr}{\widetilde{\boldsymbol{\theta}}_r}
\newcommand{\Wtett}{\widetilde{\boldsymbol{\theta}}_t}

\newcommand{\BXi}{\boldsymbol{\Xi}}
\newcommand{\BUp}{\boldsymbol{\Upsilon}}

\newcommand{\bTetar}{\boldsymbol{\Theta}_r}
\newcommand{\bTetat}{\boldsymbol{\Theta}_t}

\begin{document}
\maketitle

\begin{abstract}
We explore the potential of a simultaneously transmitting and reflecting reconfigurable intelligent surface (STAR-RIS) to enhance the performance of wireless surveillance systems. The STAR-RIS is deployed between a full-duplex (FD) multi-antenna legitimate eavesdropper ($\Ev$) and a suspicious communication pair. It reflects the suspicious signal towards the suspicious receiver ($\SD$), while simultaneously transmitting the same signal to $\Ev$ for interception purposes. Additionally, it enables the forwarding of a jamming signal from $\Ev$ to $\SD$, which is located on the back side of the STAR-RIS. To enhance the eavesdropping non-outage probability, we formulate a non-convex joint optimization problem to design the beamforming vectors at $\Ev$ and reflection/transmission phase shift matrices at the STAR-RIS. We adopt the block coordinate descent (BCD) algorithm and propose an approach, mainly based on semi-definite relaxation (SDR) and successive convex approximation (SCA), for solving the resulting decoupled sub-problems. Finally, we compare the performance of the proposed design against low-complexity zero-forcing (ZF)-based beamforming designs.

\let\thefootnote\relax\footnotetext{The work of M. Mohammadi and M. Matthaiou was
supported by the European Research Council
(ERC) under the European Union’s Horizon 2020 research
and innovation programme (grant agreement No. 101001331).}
\end{abstract}


%
\vspace{-1.4em}
\section{Introduction}\label{sec:startRIS}
Wireless information surveillance can allow authorized parties (e.g., National Security Agency, military) to legally supervise and identify abnormal user behaviors in wireless networks~\cite{Zhang:MAg:2018}.  
It involves utilizing physical layer techniques, such as jamming-assisted~\cite{Zhang:MAg:2018,Mobini:TIFS:2019} and spoofing relaying-assisted proactive eavesdropping~\cite{Zeng:JSTSP:2016}, to manipulate the suspicious link aligned with the legitimate eavesdropping channel, i.e., the link between the suspicious transmitter ($\ST$) and $\Ev$.  This ensures that $\Ev$ can intercept dubious information.

The eavesdropping performance of these systems is practically limited. Successful monitoring can only be achieved when $\Ev$ is in proximity to the $\ST$ or when a direct link exists between the $\ST$ and $\Ev$~\cite{Zhao:TWC:2023}. Therefore, wireless surveillance systems primarily employ active FD relays to simultaneously forward/overhear the suspicious signal and interfere with the suspicious link~\cite{Zeng:JSTSP:2016}.
To address the challenges of delay processing and energy consumption at active relays, surveillance systems have recently adopted the emerging RIS technology~\cite{Yao:CLET:2020,Ji:WCL:2022,Hu:TVT:2022,Zhao:TWC:2023,Hu:WCL:2023}. The RIS signal enhancement/
cancellation capabilities can be leveraged  to intelligently reflect the received suspicious signal by adjusting the phase shifts of all passive elements, turning zero delay into a reality without requiring any additional transmission power for signal forwarding~\cite{wu2021intelligent}.


Existing research contributions~\cite{Yao:CLET:2020,Ji:WCL:2022,Hu:TVT:2022,  Zhao:TWC:2023, Hu:WCL:2023} assume that the RIS can only reflect the incident signals, implying that the $\ST$ and $\SD$ must be located on the same side of the RIS. However, this geographical restriction may not always be met in practice and limits the applicability of RISs, as users are typically distributed on both sides of RISs. To overcome such limitations, the novel concept of STAR-RIS has been introduced~\cite{mu2021simultaneously,9690478}. Unlike conventional RISs, each element of a STAR-RIS can simultaneously reflect and transmit the incident signals, eliminating the need to confine their deployment to specific geographic areas and achieving full-space coverage. Moreover, since both the transmission and reflection coefficients can be designed, a STAR-RIS offers new degrees-of-freedom (DoF) to enhance the performance of wireless systems. Recently, a few initial works~\cite{hu2022analysis, HuJMC:2023} have been conducted to investigate the potential of STAR-RISs in wireless surveillance systems. The work in~\cite{HuJMC:2023} considered a half-duplex $\Ev$, while the authors in~\cite{hu2022analysis} deployed a dual-antenna FD $\Ev$, where one antenna is used to overhear the suspicious signal and the other antenna is employed to interfere with the $\SD$. Hence, the potential use of multi-antenna arrays at $\Ev$ remains unexplored. Additionally, the design of phase shifts was neglected in both~\cite{hu2022analysis,HuJMC:2023}.

We here explore a STAR-RIS-assisted wireless surveillance scenario, where a multi-antenna FD $\Ev$ aims to eavesdrop on a suspicious communication between a pair of $\ST$-$\SD$. The STAR-RIS adopts an energy-splitting protocol~\cite{mu2021simultaneously} and reflects the suspicious signal to the $\SD$, while simultaneously transmitting the suspicious signal to the $\Ev$ and forwards the jamming signal from $\Ev$ to the $\SD$.  Our main contributions are as follows:
\begin{itemize}
    \item We jointly optimize the passive transmission/reflection coefficients at the STAR-RIS and the active transmit/receive beamforming vectors at the FD $\Ev$. Although the resulting problem is a complicated non-convex optimization problem, we solve it by adopting the BCD algorithm and decompose it into tractable sub-problems. Then, we propose an approach based on SDR and SCA for solving the decoupled sub-problems. 
    \item As a benchmark, two low-complexity designs are proposed, where the ZF principle is used at the beamforming design stage to effectively cancel the self-interference (SI) at the FD $\Ev$. Our numerical results reveal that the proposed joint beamforming and phase-shift design can significantly improve the information surveillance performance of the system compared to the benchmarks.
\end{itemize}


\emph{Notations:}
We use bold upper case letters to denote matrices, and lower case letters to denote vectors. The notations $(\cdot)^{\dag}$ and   $(\cdot)^{T}$ denote the Hermitian transpose and transpose, respectively;  $\Vert\cdot\Vert$ denotes the Euclidean norm of a complex vector; $|\cdot|$ denotes the absolute value of a complex scalar;  $\trace(\cdot)$ and $(\cdot)^{-1}$ denote the trace and inverse operation of a matrix; $\diag\{\qA\}$ and $\diag\{\qa\}$ represent a vector whose elements are extracted from the main diagonal elements of the matrix $\qA$ and a diagonal matrix with $\qa$ on its main diagonal, respectively; $\qI_M$ represents the $M\times M$ identity matrix; $\boldsymbol{1}_N$ denotes an all-ones vector of size $N$.  A zero mean circular symmetric complex Gaussian variable having variance $\sigma^2$ is denoted by $\mathcal{CN}(0,\sigma^2)$.  Finally, $\mathbb{E}\{\cdot\}$ denotes the statistical expectation.

\section{System model}

\begin{figure}[t]
  \begin{center}
  \includegraphics[width=85mm, height=50mm]{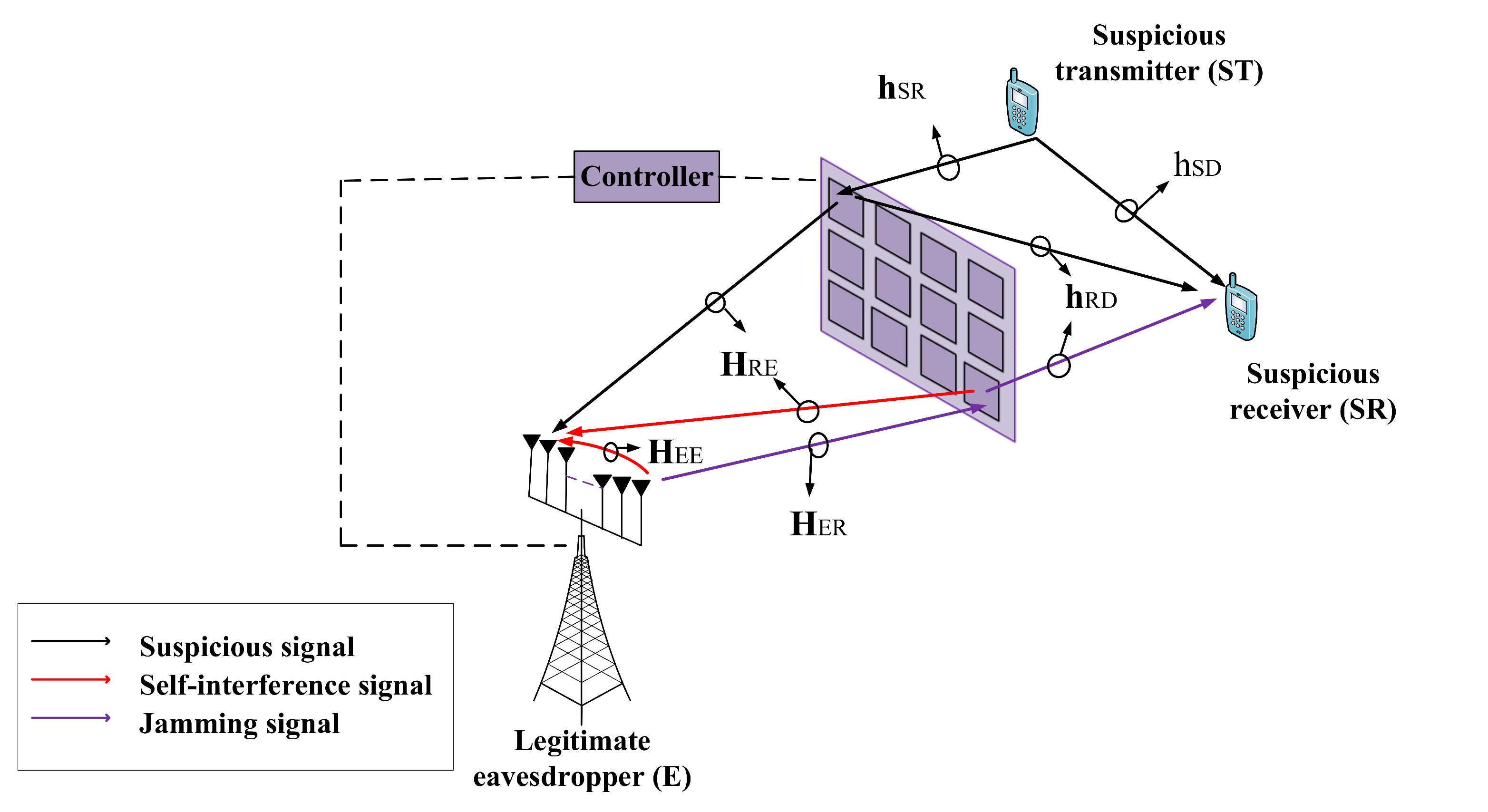}
  \vspace{0em}
  \caption{Illustration of the considered STAR-RIS-assisted wireless surveillance system.}\label{systemodel1}
  \end{center}
\end{figure}

We consider a wireless surveillance system where FD $\Ev$ aims to eavesdrop on the communication link between a single $\ST$-$\SD$ pair, as shown in Fig.~\ref{systemodel1}. However, $\Ev$ is deliberately positioned far away from the suspicious system to avoid detection, resulting in a weak eavesdropping link. To overcome this limitation, a STAR-RIS is  deployed between $\Ev$ and the suspicious system to establish an effective bridge for eavesdropping and jamming. The STAR-RIS, comprised of $N$ transmission/reflection elements, employs an energy-splitting protocol~\cite{mu2021simultaneously}. This protocol adaptively adjusts the channel power gains of the suspicious and legitimate eavesdropping links, thereby  enhancing the eavesdropping capabilities. The STAR-RIS assists $\Ev$ in overhearing the $\ST$, while simultaneously forwarding a jamming signal from $\Ev$ towards $\SD$. $\Ev$ operates in FD mode, overhearing the suspicious signal from $\ST$ via $N_{R}$ receive antennas and sending a jamming signal via $N_{T}$ transmit antennas to interfere with $\SD$. $\ST$ and $\SD$ are equipped with a single antenna each.

Let $h_{SD} = \sqrt{\gamma_{SD}}\tilde{h}_{SD}$, ${\qh_{SR}}=\sqrt{\gamma_{SR}}\tilde{\qh}_{SR}\in\mathbb{C}^{N\times 1}$, $\qh_{RD}=\sqrt{\gamma_{RD}}\tilde{\qh}_{RD}\in\mathbb{C}^{N\times 1}$, and $ \qH_{RE}=\sqrt{\gamma_{RE}}\tilde{\qH}_{RE}\in\mathbb{C}^{N_{R} \times N }$ represent the channel coefficients for $\ST$-to-$\SD$, the $\ST$-to-STAR-RIS, the STAR-RIS-to-$\SD$, and the STAR-RIS-to-$\Ev$ link, respectively, where  $\gamma_{XY}$ denotes the large-scale fading between node $X\in\{S,R\}$  and $Y\in\{D,R, E\}$. Moreover, $\tilde{h}_{SD}$, $\tilde{\qh}_{SR}$, $\tilde{\qh}_{RD}$, and $\tilde{\qH}_{RE}$  denote the small-scale fading components each having  $\mathcal{CN}(0,1)$ elements. The SI link at $\Ev$ is denoted by $ {\qH_{EE}}\in\mathbb{C}^{N_{R}\times N_{T}}$, whose elements can be modeled as $\mathcal{CN}(0,\sigma_{\mathtt{SI}}^2)$ random variables~\cite{mohammadi2016throughput}. 
We express the transmission and reflection coefficient matrices of the STAR-RIS as $\bTetat = \diag \big( \big[\sqrt{\beta_{1}^{t}}e^{j\theta_{1}^{t}},\ldots,\sqrt{\beta_{N}^{t}}e^{j\theta_{N}^{t}}\big] \big)$ and $\bTetar = \diag\big(\big[\sqrt{\beta_{1}^{r}}e^{j\theta_{1}^{r}},\ldots,\sqrt{\beta_{N}^{r}}e^{j\theta_{N}^{r}}\big] \big)$ respectively, where $\beta_{n}^{t}$ ($\beta_{n}^{r}$) and $\theta_{n}^{t}$ ($\theta_{n}^{r}$), $\forall n \in \left\{1,2,\ldots,N  \right\}$ denote the transmission (reflection) energy splitting factors and phase shifts, respectively. In addition, we note that the phase shifts $\theta_{n}^{t},~\theta_{n}^{r} \in \left[ 0, 2\pi \right]$ are generally chosen independently of each other. In contrast, according to the law of energy conservation, $\beta_{n}^{t},~ \beta_{n}^{r}\in \left[ 0,1 \right] $ are coupled with each other, i.e., they should satisfy the constraint of $\beta_{n}^{t} + \beta_{n}^{r} = 1,  \forall n \in \left\{1,2,\ldots,N  \right\}$~\cite{9690478}. Similar to~\cite{hu2022analysis,HuJMC:2023}, we adopt the same energy splitting ratio for each element of the STAR-RIS, i.e., $\beta_{1}^{t} =   \ldots  =  \beta_{N}^{t} = \beta^{t} $, $\beta_{1}^{r} =  \ldots  =  \beta_{N}^{r} = \beta^{r}$ and $\beta^{t} + \beta^{r} = 1$. 
\vspace{-1em}
\subsection{Transmission Protocol}
Assume that the $\ST$ transmits signal ${x}_s$ to $\SD$ with transmit power ${P}_s$. Hence, the received signal of the $\SD$ is given by
\begin{align}~\label{eq:YD1}
{y}_{D}\! =\!&\sqrt{\roE}\hRD^\dag\bTetat{\HHER}\wt{{x}_j}\nonumber\\
&\hspace{3em}+\sqrt{{ \Ps }}\big({h_{SD}}
+\!\hRD^\dag\bTetar\hSR \big){{x}_s}\!+\! {{n}_{D}},
\end{align}
where $\wt\in\mathbb{C}^{N\times 1}$ is the transmit beamforming vector,  $\roE$ denotes the jamming power of $\Ev$, ${x}_j$ denotes the jamming signal of $\Ev$, and $n_{D}\sim\mathcal{CN}(0, \SnD)$. Hence, the received signal-to-interference-plus-noise ratio (SINR) at $\SD$ is given by
\vspace{-0.2em}
\begin{equation}~\label{eq:gaD1}
\SINRD= \frac{{{ \Ps }}|{h_{SD}}+\hRD^\dag\bTetar\hSR|^{2} }{{\roE}|\hRD^\dag\bTetat{\HHER}\qw_t|^{2}+ \SnD }.
\end{equation}

Let $\wrr\in\mathbb{C}^{N\times 1}$  denote the receive beamforming vector at $\Ev$. Then, the received signal at $\Ev$ is given by
\vspace{-0.1em}
\begin{align}~\label{eq:YE1}
y_{E} =&\sqrt\Ps\qw_r^{\dag}\HHRE\bTetat\hSR{x}_s+\sqrt{\roE}
\big(\qw_r^{\dag}\HHEE\qw_t
\nonumber\\
&
+\qW_r^{\dag}
\HHRE\bTetar\HHER \qw_t\big){x}_j+ 
\qw_r^{\dag}{\qn}_{E}.
\end{align}
where $\qn_{E}\in\mathbb{C}^{N\times 1}$ is the received noise vector such that $\qn_{E}\sim\mathcal{CN}(\boldsymbol{0}, \SnE\qI_N)$. 
Accordingly, the received SINR at $\Ev$ can be expressed as
\begin{align}~\label{eq:PD4}
\SINRE&= \frac{\Ps|\wrr^{\dag}{\HHRE}\bTetat\hSR|^{2} }{{\roE}|\wrr^{\dag}{\HHEE}{\wt}+{\wrr^{\dag}\HHRE}\bTetar{\HHER}{\wt}|^{2}+ \SnE}.
\end{align}

As the performance metric for the considered surveillance system, we focus on the eavesdropping non-outage probability, denoted as $\NOP \triangleq\Ex\{X\}$. The indicator function $X$ in $\Ex\{X\}$ denotes the event of successful eavesdropping at $\Ev$, given by
\begin{align} 
X = \begin{cases} \displaystyle 1 & \text {if } \SINRE\geq \mathrm \SINRD,\\ \displaystyle 0 & \text {otherwise}. \displaystyle \end{cases}
\end{align}
where $X=1$ and $X=0$ indicate the eavesdropping non-outage and outage events, respectively. In other words, to achieve a reliable detection at $\SD$, $\ST$ varies its transmission rate according to $\SINRD$. Hence, if $\SINRE\geq \mathrm \SINRD$, then $\Ev$ can also reliably decode the information intended to $\SD$. On the other hand, if $\SINRE< \SINRD$, $\Ev$ is unable to decode this information without any error~\cite{Zhong:TWC:2017,Feizi:TCOM:2020}. The eavesdropping non-outage probability is mathematically represented by
\begin{align}~\label{eq:YR0}
&\NOP (\qw_r, \qw_t, \bTetar, \bTetat) = \Pr\big(\SINRE > \SINRD\big),
\nonumber\\
&\hspace{2em}=\Pr\bigg( \frac{\Ps|\wrr^{\dag}{\HHRE}\bTetat\hSR|^{2} }{{\roE}|\wrr^{\dag}{\HHEE}{\wt}+{\wrr^{\dag}\HHRE}\bTetar{\HHER}{\wt}|^{2}+ \SnE}
\nonumber\\
&\hspace{6em}>\frac{{{ \Ps }}|{h_{SD}}+\hRD^\dag\bTetar\hSR|^{2} }{{\roE}|\hRD^\dag\bTetat{\HHER}\wt|^{2}+ \SnD } \bigg).
\end{align}

\vspace{-1em}
\subsection{Problem Formulation}
The eavesdropping non-outage probability expression in~\eqref{eq:YR0} depends on the transmit/receive beamforming vectors at $\Ev$, i.e., $\qw_t$, $\qw_r$, as well as transmission/reflection matrices at the STAR-RIS, i.e., $\bTetat$, and $\bTetar$. Therefore, we can jointly optimize $\qw_t$, $\qw_r$, $\bTetat$, and $\bTetar$ to enhance the surveillance performance. To characterize the fundamental information theoretic performance limits of the considered wireless information surveillance system,  we assume that $\Ev$ has perfect channel state information (CSI) of all links~\cite{hu2022analysis,HuJMC:2023}. In practice, $\Ev$ can overhear the pilot signals sent by $\ST$ and $\SD$ to acquire the CSI of $\qh_{SR}$, $\qh_{SR}$, and $\qH_{RE}$. To this end, the proposed channel estimation method in~\cite{Wu:COML:2022} can be applied.  On the other hand, $\Ev$ can obtain the CSI of $h_{SD}$ by eavesdropping the feedback channels of the suspicious
transmitter-receiver pair~\cite{Zhang:TWC:2020}.  

Noticing that the STAR-RIS operates in energy-splitting mode, the joint optimization problem can be formulated as
\vspace{-0.1em}
\begin{subequations}\label{P2:max of subtract1}
\begin{align}
\underset{\qw_r, \qw_t, \bTetar, \bTetat}{\max}\,\, \hspace{0.5em}&
		 \NOP (\qw_r, \qw_t, \bTetar, \bTetat),
		\\
		\mathrm{s.t.} \,\,
		\hspace{2em}&
  \beta^{r}+\beta^{t} = 1,~\label{P1:beta:const}\\
		&~
   \Vert{e^{j\theta_{t}}}\Vert=\Vert{e^{j\theta_{r}}}\Vert=1,\label{P1:phase1}
   \\
		&~
   \Vert\wrr\Vert=\Vert{\wt}\Vert=1.~\label{P1:wrwt1}
\end{align}
\end{subequations}

\section{Joint Beamforming and Phase-shift Design}
In this section, we propose an optimal design to jointly optimize the beamforming vectors at $\Ev$ and transmission/reflection coefficients at STAR-RIS. 
Before proceeding, by invoking~\eqref{eq:gaD1} and~\eqref{eq:PD4}, we first re-express the objective function in a more tractable form. Therefore, the optimization problem~\eqref{P2:max of subtract1} can be recast as
\begin{subequations}\label{P2:min of subtract2}
\begin{align}~\label{eq:YR1}
\underset{\qw_r, \qw_t,  \bTetar, \bTetat}{\min}\,\, \hspace{0.1em}&\bigg( \frac{|{h_{SD}}+\hRD^\dag\bTetar\hSR|^{2} }{{\roE}|\hRD^\dag\bTetat{\HHER}\wt|^{2}+ \SnD }
\nonumber\\
&\hspace{-3em}-\frac{|\wrr^{\dag}{\HHRE}\bTetat\hSR|^{2} }{{\roE}|\wrr^{\dag}{\HHEE}{\wt}+{\wrr^{\dag}\HHRE}\bTetar{\HHER}{\wt}|^{2}+ \SnE}\bigg),
\\
		\mathrm{s.t.} \,\,
		\hspace{2em}& ~\eqref{P1:beta:const}-\eqref{P1:wrwt1}.
\end{align}
\end{subequations}
It is obvious that~\eqref{P2:min of subtract2} is a non-convex problem due to its non-convex objective and the phase shift-related constraints. To tackle this issue, we apply the widely used classic BCD algorithm to solve~\eqref{P2:min of subtract2}.  To this end, we partition the optimization variables into two
blocks, 1) receive/transmit semaphores at $\Ev$, i.e., $\wrr$ and $\wt$;  2) Transmission/reflection phase shifts at
the STAR-RIS, i.e., $\bTetat$ and $\bTetar$. Then, we minimize the objective function in~\eqref{P2:min of subtract2} by iteratively optimizing each of the above two blocks in one iteration while the other block is fixed, until convergence is reached.

\subsection{Beamforming Design at $\Ev$}~\label{sec:Ev}
By inspecting the objective function~\eqref{eq:YR1}, we find out that only the second term, i.e., $\frac{|\wrr^{\dag}{\HHRE}\bTetat\hSR|^{2} }{{\roE}|\wrr^{\dag}{\HHEE}{\wt}+{\wrr^{\dag}\HHRE}\bTetar{\HHER}{\wt}|^{2}+ \SnE} $ depends on $\qw_r$. Therefore, for a given $\qw_t$, the optimal $\qw_r$ is the solution of the following optimization problem
\begin{align}~\label{eq:wr1}
\max_{ \Vert\wrr\Vert= 1}\qquad \frac{|\wrr^{\dag}{\HHRE}\bTetat\hSR|^{2} }{{\roE}|\wrr^{\dag}{\HHEE}{\wt}+{\wrr^{\dag}\HHRE}\bTetar{\HHER}{\wt}|^{2}+ \SnE}.
\end{align}
Since~\eqref{eq:wr1} is a generalized Rayleigh ratio problem~\cite{mohammadi2016throughput}, the optimal receive beamformer can be obtained in closed-form 
\begin{align}
    \qw_r^{\ast}=\frac{(\rho_{e}\qB+\qI_{N_{R}})^{-1}{\HHRE}\bTetat\hSR }{\Vert(\rho_{e}\qB+\qI_{N_{R}})^{-1}\HHRE\bTetat\hSR\Vert},
\end{align}
where $\rho_{e}\!=\!{\roE}/{\SnE}$ is the normalized signal-to-noise ratio (SNR) of the jamming symbols, while 
$\qB\triangleq(\HHEE+\HHRE\bTetar{\HHER}
)\qW({\HHEE^{\dag}}+{\HHER^{\dag}}{\bTetar^{\dag}}
\HHRE^{\dag})$, with $\qW \triangleq{\wt}{\wt^{\dag}}$. 
By substituting $\qw_r^{\ast}$ into~\eqref{P2:min of subtract2}, the latter can be written as
\vspace{-0.1em}
\begin{align}\label{P3:min of add44}
\underset{\Vert{\wt}\Vert=1}{\min}\,\, &\hspace{2em}~
\frac{\frac{\SnE}{\SnD}|{h_{SD}}+\hRD^\dag\bTetar\hSR|^{2} }{1+\frac{{\roE}}{\SnD}|\hRD^\dag\bTetat{\HHER}\wt|^{2} }
\\
&\hspace{-2em}+\frac{\rho_{e}|{\textbf{h}_{SR}^{\dag}}{\bTetat^{\dag}}{\HHRE^{\dag}}({\HHEE} {{\qw}_{t}}+{\HHRE}\bTetar{\HHER {\wt})|^{2} }}{{1+\frac{{\roE}}{\SnD}{\wt^{\dag}}({\HHEE^{\dag}}+{\HHER^{\dag}}{\bTetar^{\dag}}\HHRE^{\dag}){{(\HHEE}+\HHRE}\bTetar{\HHER}
){\wt} }}.\nonumber
\end{align}
The optimization problem~\eqref{P3:min of add44} is non-convex because of the complex objective function. To proceed, we first introduce a slack variable $y= {1+\frac{{\roE}}{\SnD}\trace( \qW{\HHER^{\dag}}{\bTetat^{\dag}}{\HHRD^{\dag}} \hRD^\dag\bTetat }{\HHER} )$. Then, by employing the SDR technique~\cite{Luo:TSP:2010} to relax the quadratic terms of the beamformers in the objective function and constraints, the original problem is reformulated as follows
\vspace{-0.6em}
\begin{subequations}\label{P4:min of add55}
\begin{align}
\underset{ y, \qW\succeq0}{\min}\,\,&\hspace{1em}~
\frac{\frac{\SnE}{\SnD}|{h_{SD}}+\hRD^\dag\bTetar\hSR|^{2} }{y }
\nonumber\\
&\hspace{0em}+\frac{\rho_{e}\trace\big( \qW\big[{\textbf{U}^{\dag}}{\HHRE}\bTetat{\hSR}{ \hSR^{\dag} }\bTetat^{\dag}{\HHRE^{\dag} }\textbf{U} \big] \big) }{{1+\frac{\roE}{\SnD}\trace( \qW{\textbf{U}^{\dag}}{\textbf{U}} ) }},
\\
            \mathrm{s.t.} \,\,
		&\hspace{0.5em}~
             \ y= {1+\frac{{\roE}}{\SnD}\trace( \qW{\HHER^{\dag}}{\bTetat^{\dag}}{\hRD} \hRD^\dag\bTetat }{\HHER} ),~\label{P3:wrwt1}
          \\
		 &\hspace{0.5em}~
    \ \trace(\qW)=1,\label{P3:phase1}  
\end{align}
\end{subequations}
where $\textbf{U} \triangleq {\HHEE}+{\HHRE}\bTetar{\HHER}$.
Problem ~\eqref{P4:min of add55} is still non-convex in the objective function and constraints ~\eqref{P3:wrwt1} and ~\eqref{P3:phase1}. 
Note that when $y$ is fixed, problem~\eqref{P4:min of add55} becomes a quasi-convex optimization problem, which can be converted into a convex SDP problem after some transformations. Hence, problem~\eqref{P4:min of add55} can be solved by the two-stage optimization procedure~\cite{Zhong:TWC:2017}, where the inner stage is an SDP problem with fixed $y$, while the outer stage is a one dimensional line search problem over $y$. In particular, the one dimensional problem is 
\vspace{-1em}
\begin{subequations}\label{P5:min of add}
\begin{align}
\underset{y}{\min}\,\, &\hspace{1em}~
f(y)+\frac{\frac{\SnE}{\SnD}|{h_{SD}}+\hRD^\dag\bTetar\hSR|^{2} }{y},
\\
            \mathrm{s.t.} \,\,
		&\hspace{2em}~
   1<y<{1+\frac{{\roE}}{\SnD}|\hRD^\dag\Theta_{t}{\HHER}\wt|^{2} },~\label{P3:wrwt}
\end{align}
\end{subequations}
where $f(y)$ is the optimal value of the inner optimization problem presented below
\begin{subequations}\label{P6:min of add5665}
\begin{align}
\underset{s>0, \textbf{Z}\succeq{\bf 0}}{\min}\,\, &\hspace{1em}~
\rho_{e}\trace\big( \textbf{Z}\big[{\textbf{U}^{\dag}}{\HHRE}\bTetat{\hSR}{ \hSR^{\dag} }\bTetat^{\dag}{\HHRE^{\dag} }\textbf{U} \big] \big),
\\
            \mathrm{s.t.} \,\,
		&\hspace{0.5em}~
    \ \trace(\textbf{Z})=s,\label{P3:phase}
 \\
		 &\hspace{0.5em}~
    \ {{s+\frac{{\roE}}{{\SnD}}\trace( \textbf{Z}
    {\textbf{U}^{\dag}}{\textbf{U}} ) }}=1,\label{P3:phase}    
\\
		 &\hspace{0.5em}~
    \ {s}{(y\!-1)}\!=\!{\frac{{\roE}}{\SnD}\trace( \textbf{Z}{\HHER^{\dag}}{\bTetat^{\dag}}{\hRD} \hRD^\dag\bTetat }{\HHER} ),\label{P3:phase}      
\end{align}
\end{subequations}
where we have introduced ${\textbf{Z} = s\qW}$, while ${s > 0}$ satisfies ${{{s+\frac{{\roE}}{{\SnE}}\trace( s\qW{\textbf{U}^{\dag}}{\textbf{U}} ) }}=1}$. Problem~\eqref{P6:min of add5665} consists of a linear objective function with a set of linear constraints, hence it is a convex SDP problem that can be efficiently solved. Thus, optimization problems~\eqref{P5:min of add} and~\eqref{P6:min of add5665} are iteratively solved to provide the optimal transmit beamformer, $\wt^\star$.

\subsection{STAR-RIS Transmission/Reflecting Phase Design }~\label{sec:RIS}
In this subsection, we propose a joint design of $\bTetar$ and $\bTetat$ at the STAR-RIS for given $\wt$ and $\wrr$ at $\Ev$.
To this end, the optimization problem~\eqref{P2:min of subtract2} is formulated as
\begin{subequations}\label{P7:min of add1}
\begin{align}
\underset{\bTetar, \bTetat}{\min}\,\, 
&\Big( \frac{|{h_{SD}}+\hRD^\dag\bTetar\hSR|^{2} }{{\roE}|\hRD^\dag\bTetat{\HHER}\wt|^{2}+ \SnD }
\nonumber\\
&\hspace{0em}-\frac{|\wrr^{\dag}{\HHRE}\bTetat\hSR|^{2} }{{\roE}|\wrr^{\dag}{\HHEE}{\wt}+{\wrr^{\dag}\HHRE}\bTetar{\HHER}{\wt}|^{2}+ \SnE}\Big),
\label{P7:c1}\\
		\mathrm{s.t.} \,\,
		&\hspace{2em} ~\eqref{P1:beta:const},\eqref{P1:phase1}.
\end{align}
\end{subequations}

By defining ${\Wtetr} = \diag(\bTetar)^{\dag}$, ${\Wtett} =\diag(\bTetat)^{\dag}$, 
$\qa_{k_{1}} = \diag(\hRD^\dag){\hSR}$, 
${\overline\qa_{k_{1}}}=[\qa_{k_{1}}^{H}, h_{SD}^{H}]^{H}$, $\qA_{k_{1}}=  {\overline\qa_{k_{1}}}{\overline\qa_{k_{1}}^{H}}$, $\qa_{k_{2}} = \diag(\hRD^\dag){\HHER}{\wt}$, ${\overline\qa_{k_{2}}}=[\qa_{k_{2}}^{H},  0]^{H}$, $\qA_{k_{2}}={\overline\qa_{k_{2}}}{\overline\qa_{k_{2}}^{H}}$, 
$a_{k_3} = \diag(\wrr^{\dag}{\HHRE}){\hSR}$, ${\overline\qa_{k_{3}}}=[\qa_{k_{3}}^{H},  0]^{H}$,  $\qA_{k_3} ={\overline\qa_{k_{3}}}{\overline\qa_{k_{3}}^{H}}$, 
 $a_{k_{4}} = \diag(\wrr^{\dag}{\HHRE}){\HHER}{\wt}$, 
 $b = \wrr^{\dag}{\HHEE}{\wt}$, ${\overline\qa_{k_{4}}}=[\qa_{k_{4}}^{H},  b]^{H}$,  $\qA_{k_{4}} =\overline\qa_{k_4}{\overline\qa_{k_{4}}^{H}}$, $\Otetr= [\Wtetr,  1]$, 
 ${\Otett}= [\Wtett,  1]$, $\qQ_{r} = \Otetr^{H}\Otetr$, and $\qQ_{t} = \Otett^{H}\Otett$, the optimization problem~\eqref{P7:min of add1}  can be reformulated as 
\begin{subequations}\label{P7:min of add2121}
\begin{align}
\underset{ \qQ_{t},\qQ_{r}\succeq0}{\min}\,\, &\hspace{1em}~
\frac{\trace(\qA_{k_{3}}\qQ_{t}) \trace(\qA_{k_{2}}\qQ_{t}) }
{\trace(\qA_{k_{4}}\qQ_{r}) \trace(\qA_{k_{1}}\qQ_{r})},~\label{eq:p77:ob}
\\
		\mathrm{s.t.} \,\,
		 &\hspace{2em}  \Diag\{\qQ_{r}\}+\Diag\{\qQ_{t}\}=\boldsymbol{1}_N,~\label{eq:p77:d}\\
 &\hspace{2em} \Rank(\qQ_{r})=1,~ ~\label{eq:p77:h}\\
&\hspace{2em} \Rank(\qQ_{t})=1. ~\label{eq:p77:j}
\end{align}
\end{subequations}

The optimization problem~\eqref{P7:min of add2121} is still non-convex due to non-convex objective function and constraints~\eqref{eq:p77:j}. To deal with the non-convex objective function, we introduce two positive slack variables $\frac{1}{I_k}= \trace(\qA_{k_{3}}\qQ_{t}) \trace(\qA_{k_{2}}\qQ_{t})$ and $S_k= \trace(\qA_{k_{4}}\qQ_{r}) \trace(\qA_{k_{1}}\qQ_{r})$. As a result, the problem~\eqref{P7:min of add2121} is recast as
\vspace{-0.3em}
\begin{subequations}\label{P7:min of add23}
\begin{align}
&\underset{ \qQ_{t},\qQ_{r}, I_k, S_k}{\min}\,\, \hspace{1em}~
\frac{1}
{I_kS_k},
\\	
&\hspace{2.5em}             \mathrm{s.t.} \,\,
		   \hspace{3em} \frac{1}{I_k} \leq \trace(\qA_{k_{3}}\qQ_{t}) \trace(\qA_{k_{2}}\qQ_{t}),~\label{P8:tr1} \\
              &\hspace{7.2em} S_k\geq \trace(\qA_{k_{4}}\qQ_{r}) \trace(\qA_{k_{1}}\qQ_{r}),~\label{P8:tr2}\\
             &\hspace{7.2em}~\eqref{eq:p77:d}-\eqref{eq:p77:j}.
\end{align}
\end{subequations}
Problem~\eqref{P7:min of add23} is still non-convex due to the first two constraints and the non-convex rank-one constraints. To deal with the non-convex constraints~\eqref{P8:tr1} and~\eqref{P8:tr2}, we apply the following lower bounds
\vspace{-0.4em}
\begin{align}
    \label{xy:ub}
    & 4xy \! \leq \! [(x\!+\!y)^2\!-\!2(x^{(n)}\!-\!y^{(n)})(x\!-\!y) 
    \!+\! (x^{(n)}\!-\!y^{(n)})^2],
    \\
    \label{minusxy:ub}
    &\hspace{-1.5em}~ -4xy \! \leq \!  [(x\!-\!y)^2\!-\!2(x^{(n)}\!+\!y^{(n)})(x\!+\!y)
    \!+\! (x^{(n)}\!+\!y^{(n)})^2],
\end{align}
where $x\geq0, y\geq0$. Then, by  relaxing the rank-one constraints for $\qQ_{t}$ and $\qQ_{r}$, we can solve the following SDP problem
\begin{subequations}\label{P7:min of add11}
\begin{align}
&\underset{ \qQ_{t},  \qQ_{r}, I_k, S_k}{\min}\,\, \hspace{0.1em}
\frac{1}{I_k S_k},
\\
&\hspace{2em} \mathrm{s.t.} \,\,\hspace{1.5em}
\!\big(\trace(\qA_{k_{3}}\qQ_{t})\!-\!\trace(\qA_{k_{2}}\qQ_{t})\big)^{\!2}\!\!\!-\!2\big(\trace(\qA_{k_{3}}\qQ_{t}^{\!(n)})
\nonumber\\
&\hspace{5em}+\trace(\qA_{k_{2}}\qQ_{t}^{(n)})\big)\!\big(\trace(\qA_{k_{3}}
\qQ_{t}) \!+\!\trace(\qA_{k_{2}}\qQ_{t})\big)
\nonumber\\
&\hspace{5em}
+ \!\big(\trace(\qA_{k_{3}}\qQ_{t}^{(n)})+\trace(\qA_{k_{2}}\qQ_{t}^{(n)})\big)^{\!2} + \frac{4}{I_k}\!\leq\!0,~\label{eq:99:b}\\
&\hspace{5em}
\big(\trace(\qA_{k_{4}}\qQ_{r})+\trace(\qA_{k_{1}}\qQ_{r})\big)^2
-2\big(\trace(\qA_{k_{4}}\qQ_{r}^{(n)})\nonumber\\       
&\hspace{5em}-
\trace(\qA_{k_{1}}\qQ_{r}^{(n)})\big)
\big(\trace(\qA_{k_{4}}\qQ_{r})\!-\!\trace(\qA_{k_{1}}\qQ_{r})\big) 
   \!
   \nonumber\\
&\hspace{4.8em}+\! \big(\trace(\qA_{k_{4}}\qQ_{r}^{\!(n)})\!-\!\trace(\qA_{k_{1}}\qQ_{r}^{\!(n)})\big)^{\!2} \!\!-\! 4S_k \!\leq\! 0,~\label{eq:p99:c1} \\ 
             &\hspace{5em}~\eqref{eq:p77:d},\eqref{eq:p77:j}. 
\end{align}
\end{subequations}

\begin{algorithm} \caption{Proposed Double-Layer Algorithm for
Solving Problem~\eqref{P7:min of add2121}}
\label{alg:Alg1}
\begin{algorithmic}[1]
\State Initialize feasible points $\left\{ {\qQ}^{\left( 0 \right)}_{t}, {\qQ}^{\left( 0 \right)}_{r}, {\qw}^{\left( 0 \right)}_{t} \right\}$.
\State {Set iteration index $n = 0$.}
\State \textbf{Repeat} 
\State For given $\qQ^{\left( 0 \right)}_{t}$ and ${\qQ}^{\left( 0 \right)}_{r}$,  solve the relaxed  problem~\eqref{P6:min of add5665} and return $\qw^{\left( n \right)}_{t}$.
\State For given $\qw^{\left( n \right)}_{t}$,  solve the relaxed problem~\eqref{P7:min of add11} and return ${\qQ}^{\left( n \right)}_{t}$ and ${\qQ}^{\left( n \right)}_{r}$.
\State Update $n = n + 1$.
\State \textbf{Until} 
the fractional decrease of the objective function value is below a predefined threshold $\epsilon_{1} > 0$ or the maximum number of inner iterations $n_{max}$ is reached.
\State Return $\qQ^{\star}_{t}$, $\qQ^{\star}_{r}$, and $\qw^{\star}_{t}$ with the current solutions $\qQ^{\left(n \right)}_{t}$, $\qQ^{\left(n \right)}_{r}$, and $\qw^{\left( n \right)}_{t}$.
\end{algorithmic}
\end{algorithm}
\setlength{\textfloatsep}{0.7cm}

This SDP problem can be solved efficiently via CVX~\cite{cvx}. The optimal solution of $\qQ_{t}$ and $\qQ_{r}$, however,
are not generally rank-one matrices. Therefore, after obtaining the optimal $\qQ_{t}$ and $\qQ_{r}$, we need to find a rank-one solution by using the Gaussian randomization procedure~\cite{Luo:TSP:2010}.

\vspace{-0.5em}
\subsection{Overall Algorithm and Complexity Analysis}
Putting together the solution for beamforming vectors and transmission/reflection matrices presented respectively in Sections~\ref{sec:Ev} and~\ref{sec:RIS}, our proposed algorithm for maximizing the eavesdropping non-outage probability is
summarized in \textbf{Algorithm~\ref{alg:Alg1}}. Since at each iteration $n$, the proposed algorithm decreases the value of the non-outage probability, convergence of the algorithm is guaranteed.

The complexity of Algorithm~\ref{alg:Alg1} is determined by the complexity of iteratively solving the SDP problem~\eqref{P6:min of add5665} and~\eqref{P7:min of add11}.  
An SDP problem with an $a\times a$ semidefinite matrix and $b$ SDP constraints is solved with complexity $\mathcal{O}\left(\sqrt{a}\left(a^3b + a^2b^2+b^3\right)\right)$ by interior-point methods~\cite{Luo:TSP:2010}. For problem~\eqref{P6:min of add5665}, we have $a=N_T$ and $b=3$, while for problem~\eqref{P7:min of add11}  we have $a=N$ and $b=3$.

\section{Low-Complexity Designs}
The optimal design necessitates an SDP approach for beamforming design at $\Ev$, which entails high computational complexity. In light of this, we propose low-complexity suboptimal beamforming designs using the ZF principle 
and linear processing, incorporating maximum ratio transmission (MRT) and maximum ratio combining (MRC).

\vspace{-1em}
\subsection{Suboptimal Designs and Problem Formulation}
\subsubsection{ZF/MRT Beamforming Design}
With ZF/MRT beamforming design, also known as RZF, ZF beamforming is employed on the receiving side to design $\wrr$, while MRT beamforming is utilized on the transmitting side to design $\wt$. To ensure feasibility, we need to deploy at least two receive antennas at $\Ev$, i.e., ${N_{R}}>1$. Accordingly, we set $\qw_t^{\MRT} =\frac{\hRD^\dag\bTetat{\HHER}}{\Vert\hRD^\dag\bTetat{\HHER}\Vert}$, and the optimal $\wrr$, which maximizes the eavesdropping non-outage probability, is the solution of
\begin{subequations}
\begin{align}\label{eq: max}
\underset{\left\Vert {\qw_r} \right\Vert = 1}{\max}\,\, &\hspace{1em}~
{|{\qw_r^{\dag}}{\HHRE}\bTetat\hSR|^{2} }, 
\\
            \mathrm{s.t.} 
		&\hspace{1.0em}~
             {{{\qw_r^{\dag}}}\big(\HHEE+\HHRE\bTetar\HHER\big){\qw_t^{\MRT}}} = 0.         
\end{align}
\end{subequations}
By using projection matrix theory~\cite{mohammadi2016throughput}, the receive beamformer, which satisfies the condition in~\eqref{eq: max}, is given by ${\qw_r^{\ZF}} = \frac{\BXi^{\bot }{\HHRE}\bTetat\hSR}{\left\Vert\BXi^{\bot }{\HHRE}\bTetat\hSR \right\Vert}$, where $\BXi^{\bot } = \qI_{N_{R}} - \frac{\tilHHEE \qw_t^{\MRT}{(\qw_t^{\MRT})^{\dag}}\tilHHEE^{\dag}}{\Vert \tilHHEE{\qw_t^{\MRT}} \Vert^{2}}$ is the projection idempotent matrix, with $\tilHHEE = \HHEE+\HHRE\bTetar\HHER$.

Accordingly, by substituting ${\qw_r^{\ZF}}$ and ${\qw_t^{\MRT}}$ into~\eqref{eq:YR1} that completely mitigates SI, the optimization problem~\eqref{P2:max of subtract1} is reduced to
\begin{subequations}\label{P2:subopt1:Tetha}
\begin{align}
\underset{ \bTetar, \bTetat}{\min}\,\, &\hspace{0.1em}~
\frac{|{h_{SD}}\!+\!\hRD^\dag\bTetar\hSR|^{2} }{{\roE}\Vert\hRD^\dag\bTetat{\HHER}\Vert^{2}\!+\! \SnD }\!-\!\frac{|({\qw_r^{\ZF}})^{\dag}{\HHRE}\bTetat\hSR|^{2} }{ \SnE},
\\
		\mathrm{s.t.} \,\,
		&\hspace{2em} ~\eqref{P1:beta:const},\eqref{P1:phase1}.
\end{align}
\end{subequations}
\subsubsection{MRC/ZF Beamforming Design}
With MRC/ZF beamforming design, which is also abbreviated as TZF, $\wt$  is designed based on the ZF principle to cancel SI, while $\wrr$ is set according to the MRC principle to maximize the received SINR at the $\Ev$, ${\text{SINR}_E}$. Therefore, ${\qw_r^{\MRC}}$ =$\frac{{\HHRE}\bTetat{\hSR}}{\Vert{{\HHRE}\bTetat{\hSR}}\Vert}$, and the optimal $\wt$, which maximizes the eavesdropping non-outage probability, is the solution of
\vspace{-0.3em}
\begin{subequations}
\begin{align}\label{eq2: max}
\underset{\left\Vert {\qw_t} \right\Vert = 1}{\max}\,\, &\hspace{1em}~
|\hRD\bTetat\HHER\qw_t|^2 ,
\\
            \mathrm{s.t.} 
		&\hspace{1em}~
             {{(\qw_r^{\MRC})^{\dag}}\big({\HHEE}+{\HHRE}\bTetar{\HHER}\big){\qw_t}} = 0.  
\end{align}
\end{subequations}

Using projection matrix theory, the receive beamformer, which satisfies the condition in~\eqref{eq2: max}, is given by ${\qw_t^{\ZF}} = \frac{\BUp^{\bot }{\hRD\bTetat\HHER}}{\left\Vert\BUp^{\bot }{\hRD\bTetat\HHER} \right\Vert}$, where $\BUp^{\bot } = \qI_{N_{T}} - \frac{\tilHHEE^{\dag} \qw_r^{\MRC}{(\qw_r^{\MRC})^{\dag}}\tilHHEE}{\Vert \tilHHEE{\qw_r^{\MRC}} \Vert^{2}}$ with $\tilHHEE = \HHEE+\HHRE\bTetar\HHER$ is the projection idempotent. 

Accordingly, by substituting ${\qw_r^{\MRC}}$ and ${\qw_t^{\ZF}}$ into~\eqref{eq:YR1} that completely mitigates SI, the optimization problem~\eqref{P2:max of subtract1} can be written as
\vspace{-0.5em}
\begin{subequations}\label{P2:subopt2:Tetha}
\begin{align}
\underset{ \bTetar, \bTetat}{\min}\,\, &\hspace{1em}~
\frac{|{h_{SD}}+\hRD^\dag\bTetar\hSR|^{2} }{{\roE}|\hRD^\dag{\bTetat}{\HHER}{{{\qw_t^{\ZF}}}}|^{2}+ \SnD }-\frac{\Vert{\HHRE}\bTetat\hSR\Vert^{2} }{\SnE},
\\
		\mathrm{s.t.} \,\,
		&\hspace{2em} ~\eqref{P1:beta:const},\eqref{P1:phase1}.
\end{align}
\end{subequations}
\vspace{-1.5em}
\subsection{Solution }
In this subsection, we present the solution of the optimization problems~\eqref{P2:subopt1:Tetha} and~\eqref{P2:subopt2:Tetha}. Before proceeding, we define $\qa_{k_{2}}^{\cs} = \diag(\hRD^\dag){\HHER}\boldsymbol{\qf}^{\cs}$, where the superscript ``$\cs$" refers to the ``beamforming scheme" with $\cs = \{\RZF, \TZF\}$, and $\boldsymbol{\qf}^{\TZF}={{\qw_t^{\ZF}}}$, while $\boldsymbol{\qf}^{\RZF}=\boldsymbol{1}_N$. Moreover, let  $\overline\qa_{k_{2}}^{\cs}=[(\qa_{k_{2}}^{\cs})^{H},  0]^{H}$, $\qA_{k_{2}}^{\cs}=\overline\qa_{k_{2}}^{\cs}(\overline\qa_{k_{2}}^{\cs})^{H}$, 
$\qa_{k_{3}}^{\cs} = \diag({\qg^{\cs}\HHRE}){\hSR}$ where $\qg^{\RZF}={({\qw_r^{\ZF}})^{^{\dag}} }$ and $\qg^{\TZF}=\boldsymbol{1}_N$, ${\overline\qa_{k_{3}}^{\cs}}=[\qa_{k_{3}}^{\cs},  0]^{H}$,  $\qA_{k_{3}}^{\cs}=\overline\qa_{k_{3}}^{\cs}(\overline\qa_{k_{3}}^{\cs})^{H}$. Therefore, the optimization problems~\eqref{P2:subopt1:Tetha} and~\eqref{P2:subopt2:Tetha} can be reformulated as 
\begin{subequations}\label{P7:min of add2}
\begin{align}
\underset{  \qQ_{t},\qQ_{r}\succeq0}{\min}\,\, &\hspace{1em}~
\frac{\trace(\qA_{k_{3}}^{\cs}\qQ_{t}) \trace(\qA_{k_{2}}^{\cs}\qQ_{t}) }
{ \trace(\qA_{k_{1}}\qQ_{r})},
\\
		\mathrm{s.t.} \,\,
		 &\hspace{2em}  \Diag\{\qQ_{r}\}+\Diag\{\qQ_{t}\}=\boldsymbol{1}_N,~\label{eq:p91:d}\\
&\hspace{2em} \Rank(\qQ_{r})=1,~ \Rank(\qQ_{t})=1. ~\label{eq:p91:j}
\end{align}
\end{subequations}
The optimization problem~\eqref{P7:min of add2} is non-convex due to~\eqref{eq:p77:ob}-\eqref{eq:p77:j}. To tackle this issue, we first introduce two slack variables $\frac{1}{\bar{I}_k} = \trace(\qA_{k_{3}}^{\cs}\qQ_{t}) (\trace(\qA_{k_{2}}^{\cs}\qQ_{t}))$ and $\bar{S}_k=  \trace(\qA_{k_{1}}\qQ_{r})$.

Now, following a similar approach as in the optimal case and relaxing the rank-one constraints, optimization problem~\eqref{P7:min of add2} is recast as
\begin{subequations}\label{P7:min of add}
\begin{align}
\underset{ \qQ_{t}, \qQ_{r}, \bar{I}_k, \bar{S}_k}{\min}\,\, &\hspace{0.2em}~
\frac{1}
{\bar{I}_k \bar{S}_k},
\\
\mathrm{s.t.} \,\,
&\hspace{0.2em}  \big(\trace(\qA_{k_{3}}^{\cs}\qQ_{t})\!-\!\trace(\qA_{k_{2}}\qQ_{t})\big)^{\!2}\!-\!2\big(\trace(\qA_{k_{3}}^{\cs}\qQ_{t}^{(n)})
     \nonumber\\
&\hspace{0.2em} +\trace(\qA_{k_{2}}\qQ_{t}^{(n)})\big)\big(\trace(\qA_{k_{3}}^{\cs}\qQ_{t})\!+\!\trace(\!\qA_{k_{2}}\qQ_{t})\big)
    \!
    \nonumber\\
&\hspace{0.2em} + \!\big(\trace(\qA_{k_{3}}^{\cs}\qQ_{t}^{(n)})+\trace(\qA_{k_{2}}\qQ_{t}^{(n)})\big)^{\!\!2}+ \frac{4}{\bar{I}_k}\leq\! 0,~\label{eq:p99:c2}  \\
              &\hspace{1em} \!\trace(\qA_{k_{1}}\qQ_{r}) - \bar{S}_k  \!\leq\! 0,~\label{eq:p99:c3}\\
             &\hspace{0.5em}~\eqref{eq:p91:d}.
		~\label{eq:p99:c}
\end{align}
\end{subequations}
This SDP problem can be solved efficiently via CVX~\cite{cvx}, in an iterative way. After obtaining the optimal $\qQ_{t}$ and $\qQ_{r}$, we need to find a rank-one solution by using the Gaussian randomization procedure~\cite{Luo:TSP:2010}, if the results are not rank-one.

\begin{figure}[t]
  \begin{center}
  \includegraphics[width=85mm, height=60mm]{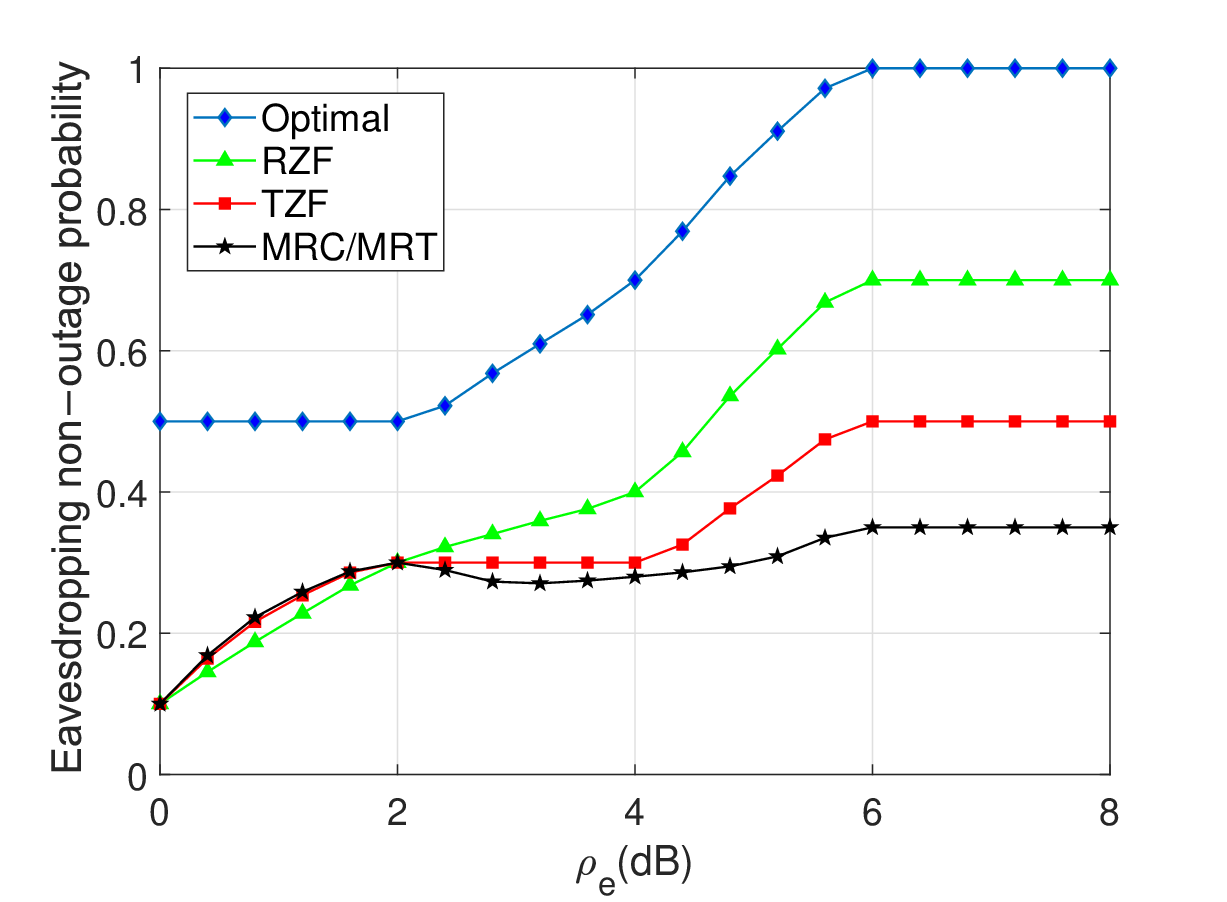}
  \caption{Eavesdropping non-outage probability for the proposed optimal and suboptimal designs versus $\rho_e$ ($N_R = N_T = 4$).
}\label{circuit_diagram1}
\vspace{-1em}
  \end{center}
\end{figure}

\section{Numerical Results}
In this section, numerical results are presented to evaluate the performance of the proposed proactive eavesdropping schemes. The optimal scheme refers to the case where the beamforming vectors at $\Ev$ and transmission/reflection coefficients at STAR-RIS are jointly optimized. Unless otherwise stated, in all the simulations, we set $\SnE \!=\! \SnD\!=\! 1$, and  $\epsilon_{1}=10^{-3}$. We  model the large-scale fading as $\gamma_{XY}= C_0(d_{XY}/D_0)^{-\mu}$, where $d_{XY}$ is the individual link distance, $C_0=-30$ dB is the reference channel gain at
a distance of $D_0=1$ m, and $\mu $ denotes the path loss exponent of the individual link  and set $\mu \!=\! 3.6$~\cite{Zhao:TWC:2023}. Moreover, the normalized SNR of the suspicious link, i.e., $\rho_s\!\!\triangleq\!\! P_s/\SnD$, is set to $10$ dB. As a benchmark, we also include the results for MRC/MRT beamforming design with optimized trans-
mission and reflection coefficient matrices at STAR-RIS.

Figure~\ref{circuit_diagram1} shows the eavesdropping non-outage probability versus $\rho_e$ for the proposed optimal and suboptimal schemes. We observe that the optimal scheme yields the best performance, reflecting the impact of joint beamforming and transmission/reflection phase shift design.  Moreover, by increasing $\rho_e$, the RZF outperforms all suboptimal designs, while MRC/MRT outperforms the RZF and TZF schemes in the low $\rho_e$ regime. The superior performance of RZF over TZF at higher values of $\rho_e$ can be attributed to the transmission of stronger jamming signals (RZF sacrifices one DoF at the transmit side to cancel SI), leading to an increased dominance of the jamming phase.  

\begin{figure}[t]
  \begin{center}
  \includegraphics[width=85mm, height=60mm]{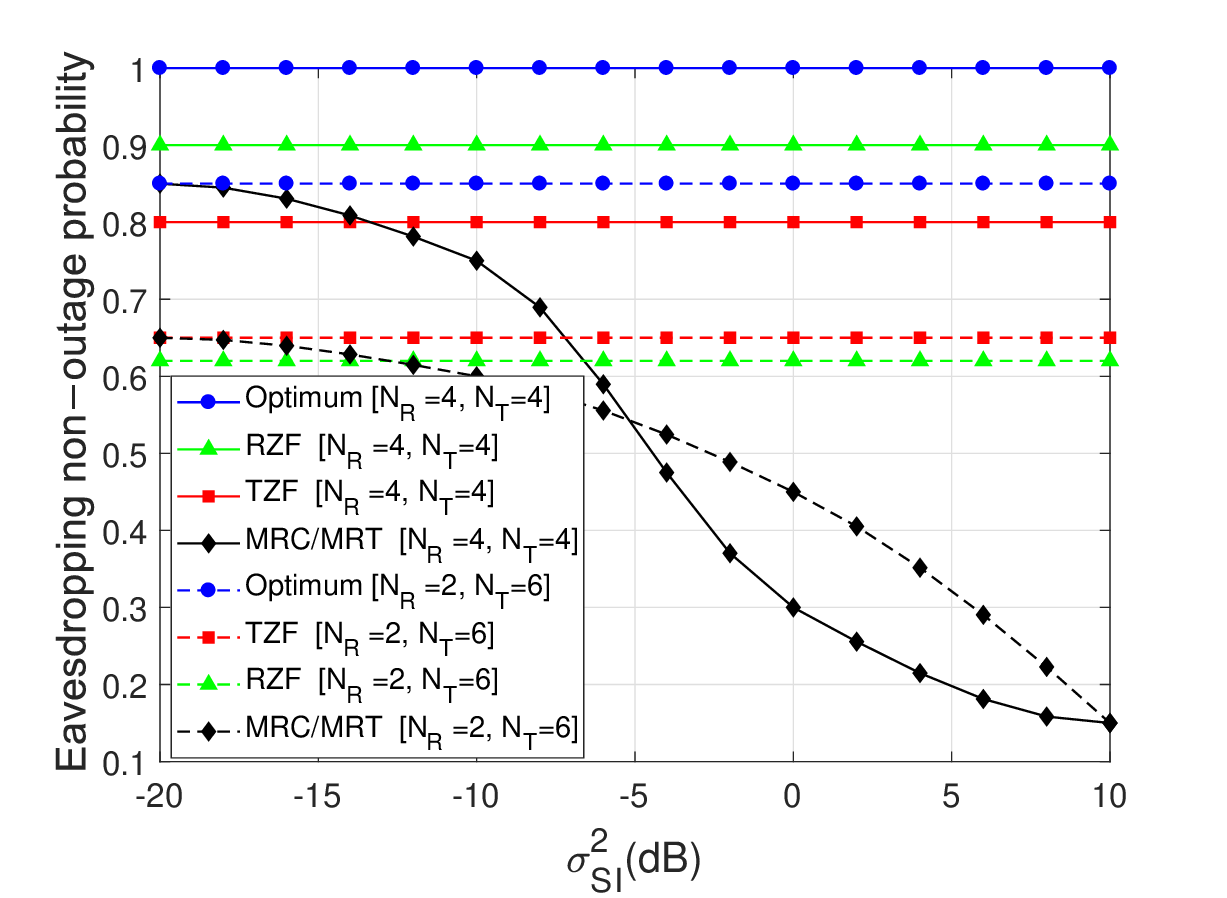}
  \caption{Eavesdropping non-outage probability 
  for the proposed optimal and suboptimal designs versus $\sigma_{\mathtt{SI}}^2$ ($\rho_e=10$ dB).
}\label{circuit_diagram2}
  \end{center}
  \vspace{-1em}
\end{figure}

Figure~\ref{circuit_diagram2} shows the eavesdropping non-outage probability versus the SI strength for the proposed schemes. We observe that the optimal design can effectively cancel SI. However, for a given number of antennas at $\Ev$, increasing the number of receiving antennas can improve performance. As expected, SI does not affect the ZF-based suboptimal schemes, while the eavesdropping non-outage probability of the MRC/MRT scheme significantly decreases when the SI strength increases. Moreover, with a decrease in the number of receiving antennas, $N_R$, the MRC/MRT scheme outperforms the ZF/MRT beamforming scheme at lower $\sigma_{\mathtt{SI}}^2$ values. This is due to the fact that decreasing $N_R$ results in weak SI, which is beneficial for the MRC/MRT design.

\section{Conclusion}
We proposed a STAR-RIS-assisted proactive eavesdropping system, where the STAR-RIS is deployed to assist a FD multi-antenna $\Ev$ in overhearing a $\ST$ and interfering with a $\SD$ simultaneously. We formulated a non-convex joint optimization problem to design the beamforming vectors at $E$ and reflection/transmission phase shift matrices at the STAR-RIS. Moreover, low-complexity ZF-based beamforming designs were proposed that can balance between performance and system complexity. Our results suggest that the optimal design can effectively cancel the SI. Moreover, by increasing the number of receiving antennas at FD $\Ev$, the surveillance performance of the optimal design is improved.

In the future, our work will involve monitoring multiple untrusted communication links. Moreover, we can explore other operating protocols for the STAR-RIS operation in wireless surveillance systems, namely, mode switching and time switching.
\balance
\bibliographystyle{IEEEtran}
\bibliography{IEEEabrv,references}

\end{document}